\documentclass[a4paper,fleqn]{cas-dc}
\usepackage[authoryear]{natbib}
\usepackage{placeins}
\usepackage{graphicx}

\setcitestyle{aysep={,},citesep={,}}
\usepackage{tikz}
\usetikzlibrary{arrows.meta, positioning}
\def\tsc#1{\csdef{#1}{\textsc{\lowercase{#1}}\xspace}}
\tsc{WGM}
\tsc{QE}
\makeatletter

\renewcommand\subsection{\@startsection{subsection}{2}{\z@}%
  {-3.25ex\@plus -1ex \@minus -.2ex}%
  {1.5ex \@plus .2ex}%
  {\normalfont\itshape}}

\renewcommand\subsubsection{\@startsection{subsubsection}{3}{\z@}%
  {-3.25ex\@plus -1ex \@minus -.2ex}%
  {1.5ex \@plus .2ex}%
  {\normalfont\itshape}}

\makeatother

\begin{document}
\let\WriteBookmarks\relax
\shorttitle{Optical Point-Spread Function via Defocus Phase Retrieval in Scanning Electron Microscopy}    
\shortauthors{Kamal et al.}  
\title [mode = title]{Optical Point-Spread Function via Defocus Phase Retrieval in Scanning Electron Microscopy} 
\author[1]{Surya Kamal}[orcid=0000-0002-5059-2679]
\cormark[1]
\ead{surya.kamal@rit.edu}
\author[1]{Kevin Mogere Nyaburi}
\author[1]{Richard K. Hailstone}
\affiliation[1]{organization={Chester F. Carlson Center for Imaging Science, Rochester Institute of Technology},
            addressline={54 Lomb Memorial Drive}, 
            city={Rochester},
            postcode={14623}, 
            state={New York},
            country={United States}}

\cortext[1]{Corresponding author}

\begin{abstract}
The point-spread function of the probe-forming optics ($PSF_{optics}$) is reported in an uncorrected (without multipole correctors) scanning electron microscope (SEM). In an SEM, information about the electron probe is generally lost as the beam interacts with the specimen. Here, we demonstrate an approach for recovering probe phase information from reconstructed probe intensity estimates. Controlled defocus was used to acquire a focal series of SEM images of $28.5,nm$ gold ($\mathrm{Au}$) nanoparticles ($\mathrm{NPs}$) on a carbon ($\mathrm{C}$) film. These images were used to reconstruct their respective probe intensities, generating a focal series of probe intensities that served as input to the phase retrieval pipeline. Using the complete description (intensity and phase) of the electron probe wavefunction at the specimen plane, we estimate the $PSF_{optics}$ for multiple data sets at a beam energy of $E = 20,keV$. These results represent an important step in demonstrating the possibility of experimental aberration quantification in uncorrected SEMs.

\end{abstract}

\maketitle

\section{Introduction}\label{1}
In more than nine decades of its development, the scanning electron microscope
(SEM) has become an invaluable imaging instrument. Its ability to offer
surface- and subsurface-specific information coupled with relatively easier
sample preparation compared to transmission electron microscopes (TEM), makes
it a robust choice for many applications such as material characterization,
semiconductor device inspection, microchip assembly, geological sampling,
forensic analysis, biomedical sciences, electron beam lithography, etc.
\citep{2023298:28659912,2023298:28659913,2023298:28659844,2023298:28659700,2023298:28659698,2023298:28659848,2023298:28659696}.
Despite its prowess as an imaging modality, the performance of the SEM is
orders of magnitude poorer than its theoretical limits. Although there are
numerous factors such as noise, charging, and deflection fields that affect its
practical resolution~\citep{2023298:31235467}, aberrations are the
primary deteriorating factor~\citep{2023298:28659916}.

In 1936, Scherzer proved that rotationally symmetric electron lenses suffer
from positive chromatic and spherical aberration~\citep{2023298:28659919}. This initiated the quest for aberration
correction in electron optics, and after nearly six decades of research,
chromatic and spherical aberration correction was demonstrated in a low-voltage
SEM~\citep{2023298:28659920,2023298:28659923}. This was a breakthrough
in electron optics, and \citet{2023298:28659920} and \citet{2023298:28659923} achieved a practical
resolution of ${<}2~\mathrm{nm}$ at a beam energy of $E = 1~\mathrm{keV}$ by implementing the
multipole corrector design proposed by  \citet{2023298:28659925}.
Since then, other aberration-corrected SEMs have been developed
\citep{2023298:28702039,2023298:28702040,maas2025}.

Most of the research in aberration correction has been centered on perfecting
the electron optics \citep{2023298:28659934}. The idea is to combine
multipole elements with the lens field to mitigate the effect of lower-order
aberrations, as they limit the resolution. Although efficacious, multipole
aberration correctors have some inherent drawbacks. They have to be controlled
and tuned with extreme precision; the resolution is still limited by
higher-order residual aberrations and, most importantly, they are very
expensive \citep{2023298:28659936}. In many cases, the base aberration
correctors are much more costly than the SEM itself. Over time, numerous
designs have been proposed for aberration correction, including foil
correctors \citep{2023298:28660039}, electron mirrors
\citep{2023298:28660038}, electrostatic correctors
\citep{2023298:28659975,2023298:28660037,maas2025}, etc., but they are all still
focused on correcting the optics.

Aberration diagnostics are a precursor to accurate aberration correction. For
transmission electron microscopy (TEM) and scanning transmission electron
microscopy (STEM), the exit wave is imaged onto a pixelated detector. Hence,
techniques for aberration diagnostics using ronchigrams and diffractograms are
well established \citep{2023298:28659933,2023298:28659932}. In an SEM,
as the focused electron probe scans the specimen, the beam information is lost.
Therefore, no standardized aberration quantification methods exist for the
uncorrected SEM.

The aberration correction approaches reported by \citet{2023298:28659920} and \citet{2023298:28659923} rely on theoretical estimates of aberration contributions to the aberration eikonal rather than direct aberration sensing. Subsequent developments established a linear relationship between multipole field strengths and aberration coefficients using approximate probe intensities \citep{2023298:28659929} obtained from SEM images, enabling improved aberration estimation and automated correction. \citet{2023298:28659927} further demonstrated aberration quantification using a STEM camera, providing a more direct means of sensing residual optical aberrations in aberration-corrected SEMs. Nevertheless, most commercially available SEM aberration correctors remain based on the original implementation of \citet{2023298:28659920} and \citet{2023298:28659923}, with incremental upgrades. More recently, a purely electrostatic quadrupole-octupole aberration corrector improved LV-SEM resolution and contrast nearly threefold, achieving $3.0\;nm$ edge resolution at $1\;keV$ \citet{maas2025}. While these developments demonstrate effective approaches for estimating and correcting low-order aberrations, quantitative reconstruction of the coherent probe wavefront in conventional SEM has received comparatively limited attention.

Another approach to aberration correction is to accept imperfections in the
optics and introduce a reconstruction step to correct the final output (e.g.,~holography). Motivated by recent developments in dynamic electron beam
shaping \citep{2023298:28660044,2023298:28660048,2023298:28660046,2023298:28660193,2023298:28854940,2023298:28967792}
and the need for more efficient and accessible approaches to aberration
correction, we propose wavefront-sensing-based aberration correction in
SEM \citep{2023298:28660050,2023298:29047697,2023298:29047698}. Instead
of tuning multipole correctors to compensate for individual aberrations, the
idea is to accurately measure the aberrated wavefront itself, thereby enabling
the correction of all aberrations simultaneously and, in principle, the
formation of aberration-corrected probes.

As the beam information (intensity and phase) is lost during image formation, estimating the aberrated wavefront becomes a challenging task. Degradation in imaging resolution due to aberrations can be largely eliminated by an accurate measurement of the optical point-spread function of the probe-forming optics ($PSF_{optics}$). Furthermore, as we push the limits of SEM resolution, even minute features in the probe distribution become relevant. Therefore, reducing the performance of an SEM to a single probe width is unrealistic. An accurate characterization of any optical system is pivotal to understanding and improving its performance. Hence, another crucial application of measuring the $PSF_{optics}$ is establishing a pragmatic metric for evaluating the quality of the SEM optics.

In the SEM literature, the probe intensity distribution has often been referred
to as the point-spread function (PSF) of the SEM
\citep{kandel,2023298:28749914,2023298:28744420,2023298:31201338}.
While it is true that the probe intensity distribution contributes to the
blurring of specimen features during image formation, it is important to
clarify that this does not make it equivalent to a PSF. Rather, the probe
intensity serves as a blurring function. Referring to it as the PSF of the SEM
is, strictly speaking, a misnomer, one that we have also used in our earlier
publications, but which we aim to refine in this work. In this paper, we define
the optical point-spread 
function for the SEM optics
in~\eqref{dfg-666028845b46}. This distinction is critical for a precise
understanding of the results presented.

Automatic aberration correction has also been explored through image-based
optimization. \citet{2023298:28659929} introduced a method
for automatic aberration correction based on the digitization of axial
geometrical aberration coefficients. Their approach uses deconvolution of a
focused and a defocused SEM image to estimate the probe intensity distribution
and quantify third-order aberration coefficients, which are subsequently used
in a linear feedback loop for iterative hardware correction. More recently, \citet{cho2024automatic} proposed a pattern-independent beam
optimization framework based on electron beam kernel estimation. By applying
blind deconvolution to a focal series of SEM images, they jointly estimate the
beam kernel and specimen image and optimize focus, astigmatism, and aperture
alignment. While highly effective for automated microscope tuning across
diverse samples, their approach primarily serves as an optimization framework
and assumes a Gaussian beam kernel, rather than providing a wave-optical
characterization of the electron probe.

This work distinguishes itself from these previous studies by fundamentally characterizing the optical point-spread function ($PSF_{optics}$) in uncorrected SEMs without the need for expensive multipole hardware. Unlike the work of Uno \textit{et al.}, which is intrinsically tied to hardware-specific feedback loops for correction, our approach focuses on establishing a pragmatic diagnostic metric for evaluating the true quality of SEM optics. Furthermore, while Cho \textit{et al.}  rely on image-derived kernels for parameter optimization, our work employs a full wave-optical model. By using a defocus-diversity, non-interferometric phase-retrieval method on focal-series images of gold nanoparticles, we explicitly recover the lost phase information of the electron probe wavefunction. This provides a deeper physical understanding of wavefront aberrations and a complete description of the probe beyond the Gaussian approximations used in previous optimization frameworks.

In this work, we report measurements of the $PSF_{optics}$ in an SEM at a beam
energy of $E = 20~\mathrm{keV}$, based on non-interferometric phase retrieval of the
electron beam wavefunction $\psi_{sp}(x_i,y_i)$. We use a focal series of SEM images to
construct a focal series of probe intensities
\citep{kandel,2023298:28744420} and recover the lost probe phase using
an iterative phase retrieval algorithm
\citep{coene1992phase,ivanov1992phase,2023298:28660190}.

\section{Theory and Background}
Modern field-emission SEMs are equipped with highly coherent electron sources therefore a wave optical description of probe formation becomes essential\unskip~\citep{2023298:28660192}. Probe forming optics consists of the electron gun, multiple apertures and multiple lenses. The probe wavefunction at the specimen, $\psi_{sp}(x_i,y_i) $ can be expressed as a function of the virtual source distribution, coherence, aperture diffraction, and the wavefront aberrations \unskip~\citep{2023298:28659915}.

\begin{figure}
	\centering \makeatletter\IfFileExists{dcb41411-6561-4ce3-a141-58c0f79fe29a-uprloptical.png}{\includegraphics[width=\columnwidth]{dcb41411-6561-4ce3-a141-58c0f79fe29a-uprloptical.png}}{\includegraphics{dcb41411-6561-4ce3-a141-58c0f79fe29a-uprloptical.png}}
	\makeatother 
	\caption{{Optical setup for an arbitrary SEM column showing probe formation. The linearity and shift-invariance property allows the system to be represented using the exit pupil. Exit pupil is the image of the limiting aperture viewed from the specimen plane into the optics. }}
	\label{f-398175dbaf6c}
\end{figure}
For the coherent case,
\let\saveeqnno\theequation
\let\savefrac\frac
\def\dispfrac{\displaystyle\savefrac}
\begin{eqnarray}
	\let\frac\dispfrac
	\gdef\theequation{1}
	\let\theHequation\theequation
	\label{dfg-4c44c2651550}
	\begin{array}{@{}l}\begin{array}{l}\psi_{sp}(x_i,y_i)\;=h(x_i,y_i)\;\otimes\psi_g(x_i,y_i)\;\\\;\;\;\;\;\;\;\;\;\;\;\;\;\;\;\;\;\;=\iint\psi_g(\widetilde x,\;\widetilde y)\cdot h(x_i-\widetilde x,y_i-\widetilde y)d\widetilde xd\widetilde y\end{array}\end{array}
\end{eqnarray}
\global\let\theequation\saveeqnno
\addtocounter{equation}{-1}\ignorespaces 
where $\otimes $ represents convolution,  $(x_i,y_i) $ are the coordinates in the specimen plane, $\psi_g(x_i,y_i) $ is the geometrical demagnification of the virtual source wave function $\psi_s(x,y) $, where $\psi_o$ is a scaling factor. 
\let\saveeqnno\theequation
\let\savefrac\frac
\def\dispfrac{\displaystyle\savefrac}
\begin{eqnarray}
	\let\frac\dispfrac
	\gdef\theequation{2}
	\let\theHequation\theequation
	\label{dfg-5be870676a8d}
	\begin{array}{@{}l}\psi_s(x,y)=\;\psi_o\cdot e^{\textstyle\frac{-(x^{2}\;+\;y^{2})}{d_s^{2}}}\end{array}
\end{eqnarray}
\global\let\theequation\saveeqnno
\addtocounter{equation}{-1}\ignorespaces 
$h(x_i,y_i) $ is the coherent point-spread function of the optics, $PSF\;_{optics}^{coh} $, and $d_s $ is the width parameter of the virtual source.  The $PSF\;_{optics}^{coh} $ is defined as,
\begin{equation}
\label{dfg-666028845b46}
	\gdef\theequation{3}
\begin{aligned}
h(x_i,y_i)
&= \mathcal{F}\!\left\{
A(x_a,y_a)e^{-i\frac{2\pi}{\lambda}\varphi(x_a,y_a)}
\right\} \\
&= \iint_{-\infty}^{\infty}
A(x_a,y_a)e^{-i\frac{2\pi}{\lambda}\varphi(x_a,y_a)}
\\
&\qquad\times
e^{-i\frac{2\pi}{\lambda z_i}(x_ax_i+y_ay_i)}
\,dx_a\,dy_a .
\end{aligned}
\end{equation}
where $\mathcal F $ represents the Fraunhofer diffraction\unskip~\citep{2023298:28922785} of the exit plane distribution calculated at the specimen plane, $A $ is the image (projection) of the limiting aperture in the exit pupil plane. Exit pupil is defined as the image of the limiting aperture viewed from the specimen plane into the optics. $(x_a,y_a) $ are the coordinates in the exit pupil plane, $z_i $ is the distance between the exit pupil plane and the specimen plane, $\lambda $ is the relativistic electron wavelength. The exit pupil plane coordinates are related to the specimen plane as $x_a=(\lambda z_i)f_x $ , where $f_x $ is the spatial frequency coordinate in the specimen plane.  The phase $\varphi(x_a,y_a) $ in Equation~(\ref{dfg-666028845b46}) represents the wavefront aberrations present in the exit pupil which causes the final probe wavefunction to acquire a complex phase.  For the incoherent case, 
\let\saveeqnno\theequation
\let\savefrac\frac
\def\dispfrac{\displaystyle\savefrac}
\begin{eqnarray}
	\let\frac\dispfrac
	\gdef\theequation{4}
	\let\theHequation\theequation
	\label{dfg-0138e2db443f}
	\begin{array}{@{}l}\vert\psi_{sp}(x_i,y_i)\vert^{2}\;=\vert h(x_i,y_i)\vert^{2}\;\otimes\vert\psi_g(x_i,y_i)\vert^{2}\end{array}
\end{eqnarray}
\global\let\theequation\saveeqnno
\addtocounter{equation}{-1}\ignorespaces 
where $PSF_{optics}^{\;incoh}=\;\vert h(x_i,y_i)\vert^{{}^{2}} $ is the incoherent point-spread function of the optics. 
\subsection{Coherence}
Coherence is one of the most fundamental concepts of physics. General discussions on the topic are widely available in literature \unskip~\citep{2023298:28887089,2023298:28887090,2023298:28887165}. In electron optics, realistically all field-emission sources have partial coherence \unskip~\citep{2023298:28887617,2023298:28887632,2023298:28887633}. Even thermionic electron sources, which were thought to be incoherent, exhibit partial coherence \unskip~\citep{2023298:28887296}. Thus, the point-spread function for the probe-forming optics can be considered to lie on a spectrum between $PSF_{optics}^{coh}$ and $PSF_{optics}^{incoh}$ depending on the degree of source coherence. Numerous approaches have been developed to incorporate the effects of partial coherence into electron probe formation \unskip~\citep{2023298:28659915,2023298:28759602,2023298:28887703,2023298:28887857}. For partially coherent sources, the fields remain mutually coherent over a finite distance, defined as the coherence length $l_{coh}$. The spatial (transverse) coherence length at a distance $L$ is $l_{coh}^{spatial}=\lambda L/\sqrt{2}\pi d_s$, and the temporal (longitudinal) coherence length is $l_{coh}^{temporal}=2\lambda E/\Delta E$, where $\Delta E$ is the beam energy spread in eV \unskip~\citep{2023298:28887632,2023298:28887633}. Although finite energy spread primarily affects SEM probe formation through chromatic aberration, the purpose of the present discussion is to justify the use of a coherent wave-optical description of the probe under the operating conditions considered here.

For $E=20~\mathrm{keV}$ ($\lambda\approx8.5~\mathrm{pm}$), $d_s=20~\mathrm{nm}$, $\Delta E=0.7~\mathrm{eV}$, and an SEM column length of $L\approx300$--$400~\mathrm{mm}$, $l_{coh}^{spatial}\approx4.06$--$5.41~\mathrm{nm}$ and $l_{coh}^{temporal}\approx485.71~\mathrm{nm}$. Typically, SEM optics easily have a theoretical demagnification of $\sim10\times$, and for field-emission sources $d_s\le20$--$25~\mathrm{nm}$, corresponding to a geometrical probe image width of $d_g\le2$--$2.5~\mathrm{nm}$. Since the geometrically demagnified source size is smaller than the estimated spatial coherence length at the specimen plane, the illuminating probe is expected to remain spatially coherent over its extent. Therefore, our treatment of probe formation based on the coherent point-spread function $h(x_i,y_i)$ is well justified under these operating conditions.
\subsection{Probe Intensity Reconstruction}
\begin{figure}
	\centering
	\includegraphics[width=\columnwidth]{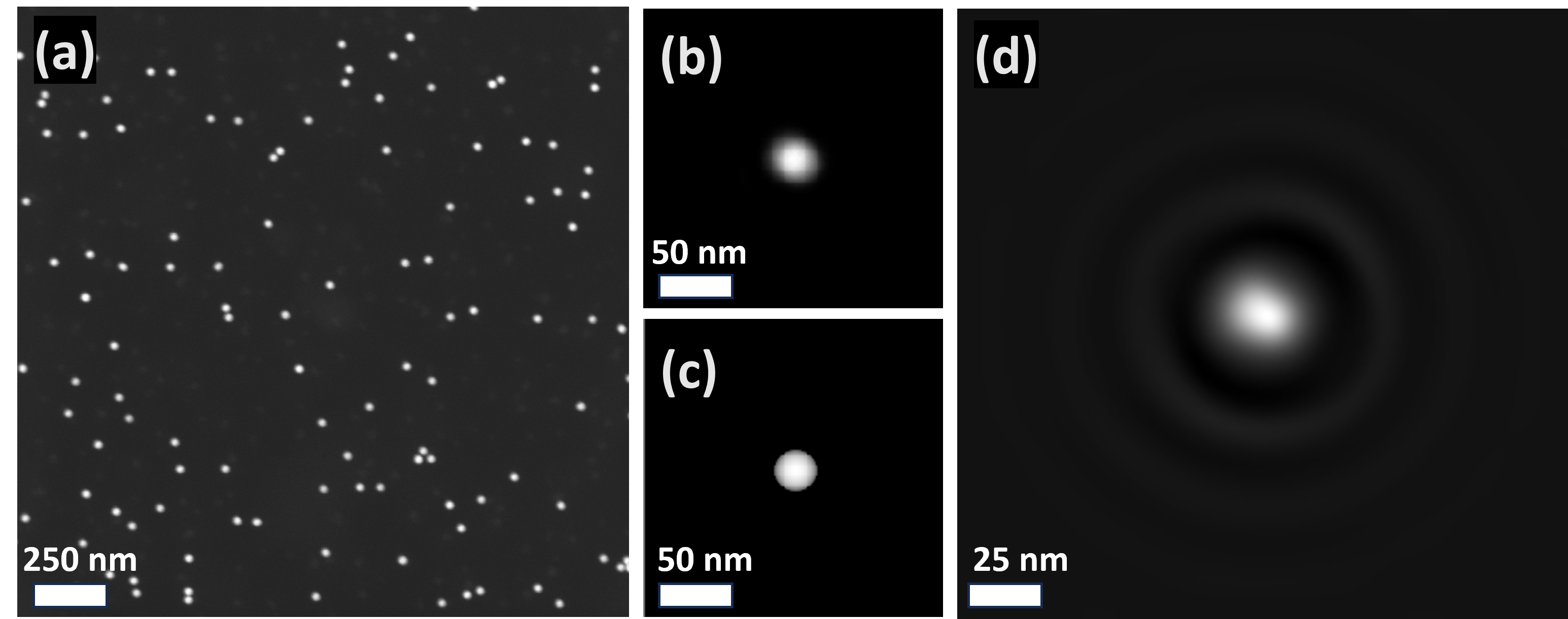}
	\caption{Probe intensity $\vert\psi_{sp}(x_i,y_i)\vert^{2}$ reconstruction process \citep{2023298:28744420}. (a) Defocused ($\Delta z\approx -2\,\mu\mathrm{m}$) BSE image of Au-NPs with multiple isolated NPs in the FOV, (b) Experimental image of a single NP captured after stacking to improve SNR, (c) Simulated image of a single NP using CASINO, (d) Reconstructed probe intensity (blurring function).}
	\label{f-6920af414af9}
\end{figure}

To calculate $h(x_i,y_i) $, we solve the inverse problem in Equation~(\ref{dfg-4c44c2651550}) and hence the complete description (intensity and phase) of probe wavefunction $\psi_{sp}(x_i,y_i) $ is needed. For reconstructing the intensity of the incident probe ($\vert\psi_{sp}(x_i,y_i)\vert^{2} $), we use the method proposed by Lifshin \unskip~\citep{kandel,2023298:28749914} and Zotta \unskip~\citep{2023298:28744420}. The process begins by capturing secondary electron (SE) or back-scattered electron (BSE) images of well dispersed gold ($\mathrm{Au} $) nanoparticles ($\mathrm{NPs} $) on a carbon ($\mathrm C $) film. Then all the individual $\mathrm{NPs} $ in the field-of-view (FOV) are segmented and stacked together to boost signal-to-noise ratio (SNR) and generate a realistic image of a single NP ($s_{real}(x_i,y_i) $). The selection of individual nanoparticles (NPs) for stacking was based on their circularity index and size. Due to the inherent $\sim5-10\%$ variation in NP size and shape, only those well within a defined threshold were included. The benefit of stacking is that it allows us to achieve a high SNR without using longer dwell times, which mitigates the chances of specimen degradation by beam damage or blurring by specimen drift. Following that, an ideal image/object structure ($s_{ideal}(x_i,y_i) $) is generated using the Monte-Carlo simulation CASINO \unskip~\citep{2023298:28749962}, which is based on the material composition, size and shape of the object, and ideal imaging conditions (aberration-free, minimum noise (Poisson-limited condition), probe width $\approx0.1\;nm $). The probe intensity is known to blur the object structure to produce the final SEM image \unskip~\citep{2023298:31201338} with additive noise ($\eta(x_i,y_i) $) present in the final image, see Equation~(\ref{dfg-052e572ae19c}).
\let\saveeqnno\theequation
\let\savefrac\frac
\def\dispfrac{\displaystyle\savefrac}
\begin{eqnarray}
	\let\frac\dispfrac
	\gdef\theequation{5}
	\let\theHequation\theequation
	\label{dfg-052e572ae19c}
	\begin{array}{@{}l}\begin{array}{l}s_{real}(x_i,y_i)\;=\left\{s_{ideal}(x_i,y_i)\;\otimes\;\vert\psi_{sp}(x_i,y_i)\vert^{2}\right\}\;\\\;\;\;\;\;\;\;\;\;\;\;\;\;\;\;\;\;\;\;\;\;\;\;+\;\eta(x_i,y_i)\end{array}\end{array}
\end{eqnarray}
\global\let\theequation\saveeqnno
\addtocounter{equation}{-1}\ignorespaces 
Noise in the image ($\eta(x_i,y_i) $) can be attributed to a combination of multiple factors like the Poisson statistics of the electron illumination, emission of the SE/BSE signals, and the signal detection process\unskip~\citep{2023298:31201341,2023298:31201340,2023298:31201339}. Since noise is always present, it is not possible to invert Equation~(\ref{dfg-052e572ae19c}) directly, therefore, a noise suppression mechanism is needed. We use a Wiener-filter\unskip~\citep{2023298:31217316} to suppress the effect of noise to estimate the probe intensity as shown in Figure~\ref{f-6920af414af9}. The choice of the defocused plane image is arbitrary and is intended solely as an illustrative example to demonstrate the probe intensity reconstruction process. It is important to note that the probe intensity is shown in a defocused plane because defocus was used as a phase diversity technique for probe phase retrieval, as explained in the \textit{Iterative Phase Retrieval} section.

\textit{} \textit{} Taking the Fourier transform of Equation~(\ref{dfg-052e572ae19c}), followed by some algebra, $\vert\psi_{sp}(x_i,y_i)\vert^{2}$ can be reconstructed using Equation~(\ref{dfg-28c710c73cfb}), where $\mathscr{F}^{-1}$ denotes the inverse Fourier transform operator, $S_{ideal}(f_x,f_y)$ and $S_{real}(f_x,f_y)$ are the spectra of the ideal and real images, respectively, and $S_{ideal}^\ast(f_x,f_y)$ is the complex conjugate of $S_{ideal}(f_x,f_y)$. The scalar $K = \tfrac{|N(f_x,f_y)|^{2}}{|\mathscr{F}\{|\psi_{sp}(x_i,y_i)|^{2}\}|^{2}}$ is defined as the ratio of the noise spectrum power to the probe intensity spectrum power. $K$ acts as a smoothing parameter and is added to the denominator to ensure numerical stability by preventing the denominator from approaching zero.
\let\saveeqnno\theequation
\let\savefrac\frac
\def\dispfrac{\displaystyle\savefrac}
\begin{eqnarray}
	\let\frac\dispfrac
	\gdef\theequation{6}
	\let\theHequation\theequation
	\label{dfg-28c710c73cfb}
	\begin{array}{@{}l}\begin{array}{l}\begin{array}{l}\vert\psi_{sp}(x_i,y_i)\vert^{2}=\\\ensuremath{\mathscr{F} }^{-1}\left[\;\left(\frac{S_{ideal}^\ast(f_x,f_y)}{\vert S_{ideal}(f_x,f_y)\vert^{2}\;+\;K}\right)\cdot S_{real}(f_x,f_y)\right]\end{array}\\\\\end{array}\end{array}
\end{eqnarray}
\global\let\theequation\saveeqnno
\addtocounter{equation}{-1}\ignorespaces

As both the noise spectrum and the probe intensity spectrum are unknown, there is no direct way of determining the optimal value of $K$. Consequently, $K$ is treated as an empirical, dataset-dependent regularization parameter. Previous studies \unskip~\citep{2023298:29003436,2023298:29003437} have systematically investigated the influence of $K$ on the reconstruction quality and demonstrated the trade-off between noise suppression and preservation of fine image features. Larger values of $K$ produce smoother reconstructions but may suppress diffraction- and aberration-related features that are essential for accurate phase recovery, whereas smaller values may amplify noise. In the present work, we selected $K\approx3162$ empirically for the experimental dataset considered, following the regularization strategy established in the above references. This value provided a stable reconstruction for the imaging conditions used in this study and should not be interpreted as a universal parameter applicable to other SEM systems or acquisition conditions.

\begin{figure*}
	\centering
	\includegraphics[width=\textwidth]{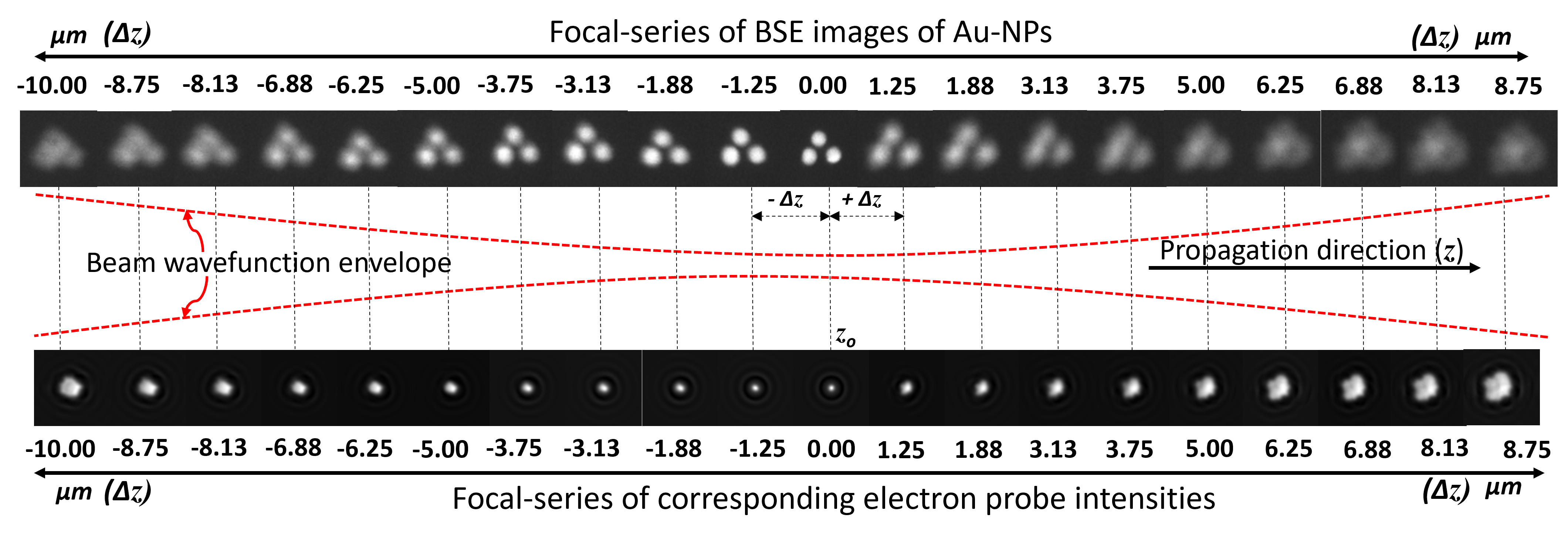}
	\caption{Data set 1: Focal series of BSE images (top) and their reconstructed probe intensities ($\vert{\psi_{sp}}_{\vert z_o\pm n\Delta z}\vert^{2}$) (bottom). $\Delta z = 1\,\mu\mathrm{m}$, 19 images were captured with a total \textit{z-length} of $\sim19\,\mu\mathrm{m}$. The actual $\Delta z$ value in $\mu\mathrm{m}$ is shown for each plane. The three NPs in the images are shown solely to illustrate the effect of defocus on the images. These are cropped segments from the full FOV.}
	\label{f-bb45751c1e8a}
\end{figure*}

\section{Experiment and Results}
\subsection{Experimental Setup}
We have used the TESCAN MIRA3 SEM with a Schottky emitter (virtual source size $\approx20\;nm $) for our experiments. All images were captured for beam energy $E=20\;keV $, FOV {\textemdash} $2.048\times2.048\;\mu m^{2} $, and number of pixels $-\;2048\times2048 $, which ensures that each image pixel corresponds to an area of $1\;nm^{2} $ on the specimen. For the experiment, the beam is focused on a central plane ($z=z_o $) and the specimen is moved along the \textit{z-axis} by $\triangle z=1\;\mu m $ steps. At every $z\;=z_o\pm\;n\cdot\triangle z\; $ ($n\;\in\;\mathbb{Z} $ and is typically around $8-10 $), a BSE image of the $\mathrm{Au}-\mathrm{NPs} $ is captured to create a through-focal-series of the specimen, as shown in Figure~\ref{f-bb45751c1e8a}.  The defocus experiment is extremely sensitive to any movement of the specimen in the \textit{z-plane}. It is important to ensure accuracy when moving the specimen by $\triangle z=1\;\mu m $ step, as most sample stages are not accurately calibrated for such a small displacement. For instance, sometimes $\triangle z=1\;\mu m $ might correspond to $\sim1.25\;\mu m $ and sometimes it might correspond to $\sim0.66\;\mu m $. Therefore, it is important to notice that difference and capture that in the iterative algorithm for accurate phase reconstruction. Once all the images are captured, then for every BSE image, the probe intensity is reconstructed using the method described above in Figure~\ref{f-6920af414af9} to construct focal-series of probe intensities $\vert{\psi_{sp}}_{\vert z_o\pm n\triangle z}\vert^{2} $. We captured three different data sets by introducing controlled astigmatism using the stigmators (see Table~\ref{tw-ad4990c395d8}). The reported stigmator percentages are included simply to illustrate that the level of astigmatism differs across the three datasets. Addition of controlled astigmastism is intentional and is done to validate its presence in the reconstructed $PSF_{optics}$ shown later in Figure~\ref{f-efbc2c1e0757}. Visualization of the first data set and the probe intensity reconstruction is shown in Figure~\ref{f-bb45751c1e8a}.

\begin{table}[!htbp]
	\caption{{Different data sets captured from the defocus experiment by introducing controlled astigmastism.} }
	\label{tw-ad4990c395d8}
	\ignorespaces 
	\centering 
	\begin{tabular}{p{\dimexpr.25\linewidth-2\tabcolsep}p{\dimexpr.21680000000000003\linewidth-2\tabcolsep}p{\dimexpr.20869999999999997\linewidth-2\tabcolsep}p{\dimexpr.3245\linewidth-2\tabcolsep}}
		data set  &  \# of images & x-stigmator & y-stigmator\\
		\hline 
		data set 1 &
		$\;\;\;\;\;\;\;\;19 $ &
		$\;\;\;\;0.4\% $ &
		$\;\;\;\;\;\;\;0.8\% $\\
		data set 2 &
		$\;\;\;\;\;\;\;\;17 $ &
		$\;\;\;\;0.3\% $ &
		$\;\;\;\;\;\;1.00\% $\\
		data set 3 &
		$\;\;\;\;\;\;\;\;17 $ &
		$\;-0.1\% $ &
		$\;\;\;\;\;\;\;0.7\% $\\
		
	\end{tabular}\par 
\end{table}
There were two qualitative observations made in the data sets. First, as we go through the focal point, we can see that the image blur and the probe intensities rotate in the orthogonal direction. It is well known that when any beam of particles is brought to a focus, it undergoes a phase shift along the longitudinal direction (along the z-axis), known as the Gouy phase shift. In the case of astigmatic beams, this also results in a rotation of the beam in an orthogonal plane \unskip~\citep{2023298:28967642}. This observed rotation therefore indicates the presence of the Gouy phase anomaly in the SEM, as we pass through the focal plane of a tightly focused astigmatic electron beam. Second, although defocus aberration is an even function, the effects of over-focus and under-focus are not the same. This implies the presence of other aberrations whose linear combination produce different effects for $-\triangle z~$and $+\triangle z~$. 

\subsection{Iterative Phase Retrieval using Defocus}
Once we have a focal-series of probe intensities, we recover the probe phase using the defocus diversity based non-interferometric phase retrieval\unskip~\citep{2023298:28660190}. The focal-series of probe intensities ($\vert{\psi_{sp}}_{\vert z_o\pm n\triangle z}\vert^{2} $) serve as the input ground truths to the iterative phase retrieval algorithm. The algorithm starts by initializing the probe wavefunction at the in-focus plane ($z=z_o $) as $\psi_{z_o}\;=\;\vert\psi_{z_o}\vert e^{i\phi_o} $, where $\vert\psi_{z_o}\vert\;=\;\sqrt{\left\vert \psi_{sp}\right\vert^{2}} $ is the reconstructed magnitude and $\phi_o $ is constant. $\psi_{z_o} $ is then propagated to the next plane $z=z_1 $ using the angular spectrum method (ASM)\unskip~\citep{2023298:28922785} and we get $\psi_{z_1}\;=\;\vert{\widetilde\psi}_{z_1}\vert e^{i\phi_1} $. Here we keep the acquired phase $\phi_1 $ and constrain the predicted magnitude $\vert{\widetilde\psi}_{z_1}\vert $ using the ground truth magnitude $\vert{\overset{}\psi}_{z_1}\vert $ according to ($\vert\psi_{z_1}\vert=(1-\gamma)\vert{\widetilde\psi}_{z_1}\vert+\gamma\vert{\overset{}\psi}_{z_1}\vert$), where $\gamma$ is the relaxation parameter\unskip~\citep{fienup1982phase}.

For this algorithm, setting $\gamma = 1$ recovers the conventional hard amplitude replacement constraint. In our experiments, $\gamma$ is typically chosen between $0.75$ and $0.95$, based on empirical tuning across multiple datasets. The updated wavefunction is propagated to the next plane. This is repeated for all planes with the reconstructed probe intensities ($\psi_{sp}\vert_{z_o\pm n\triangle z} $) serving as the ground truth until we loop back to the in-focus plane $z=z_o $, which completes one iteration. At the end of each iteration we calculate the normalized sum-squared error ($\xi $) between the ground truth ($\vert\psi_{z_o}\vert $) and predicted magnitude ($\vert{\widetilde\psi}_{z_o}\vert $) of the in-focus wavefunction shown in Equation~(\ref{dfg-7a323c2acd11}). The whole process is repeated until a stable solution is reached with a sufficiently low $\xi $.

\begin{figure}
	\centering
	\includegraphics[width=\columnwidth]{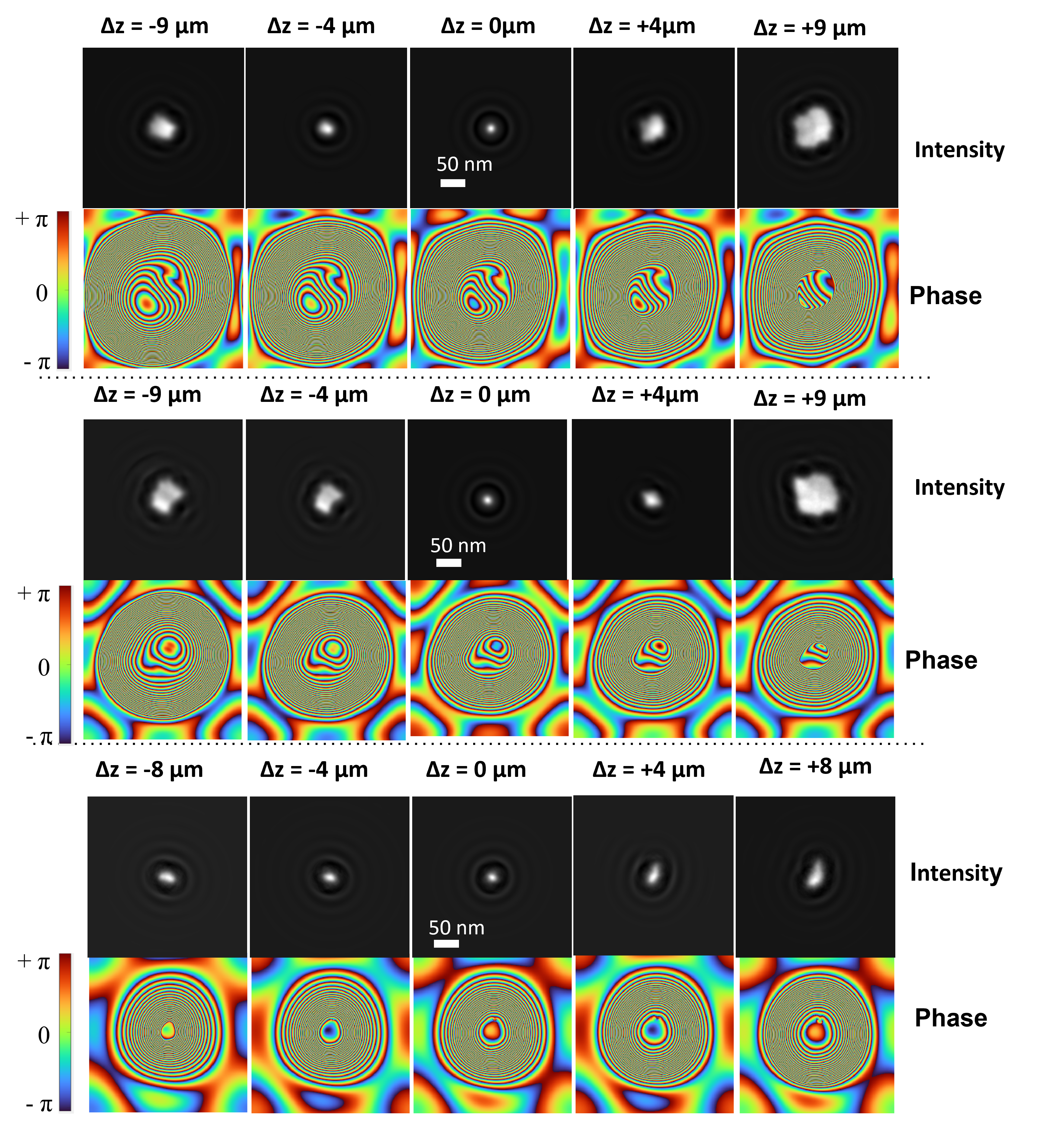}
	\caption{Phase retrieval of the electron probe using defocus. Probe intensity at different defocused planes and their corresponding recovered phases are shown. The phase is recovered for all \textit{z-planes} simultaneously. Results are shown for three different data sets with various degrees of astigmatism as shown in Table~\ref{tw-ad4990c395d8}: data set 1 (top), data set 2 (middle), and data set 3 (bottom). Data set 1---$\xi=1.82\times10^{-4}$ after $54k$ iterations; data set 2---$\xi=7.24\times10^{-5}$ after $25k$ iterations; data set 3---$\xi=1.4\times10^{-28}$ after $71k$ iterations.}
	\label{f-d563b6af9a1f}
\end{figure}

\let\saveeqnno\theequation
\let\savefrac\frac
\def\dispfrac{\displaystyle\savefrac}
\begin{eqnarray}
	\let\frac\dispfrac
	\gdef\theequation{7}
	\let\theHequation\theequation
	\label{dfg-7a323c2acd11}
	\begin{array}{@{}l}\xi\;=\frac{{\displaystyle\sum_{pixels}}\{\vert\psi_{z_o}\vert\;-\;\vert{{\widetilde\psi}_{z_o}}\vert\}^{2}}{\displaystyle\sum_{pixels}\vert\psi_{z_o}\vert^{2}}\end{array}
\end{eqnarray}
\global\let\theequation\saveeqnno
\addtocounter{equation}{-1}\ignorespaces 
All probe intensity images used for the phase retrieval pipeline were $256\times256 $ pixels with no padding. A single pixel in the probe intensity image represents an $1\;nm\times1\;nm $ area on the beam. Figure~\ref{f-d563b6af9a1f}, shows the phase retrieval of the electron probe for all the three data sets. All the reconstructed probe intensities serve as the ground truth constraint in each plane. Therefore, the phases recovered simultaneously in all the planes are the phases that would produce the probe intensities upon propagation starting at the in-focus plane ($z=z_o $). All phase maps of the estimated phases ($\widetilde\varphi\left(x_i,y_i\right) $) shown in Figure~\ref{f-d563b6af9a1f} are wrapped in the interval $\lbrack-\ensuremath{\pi},+\ensuremath{\pi}\rbrack $.

\subsection{Visualization of $PSF_{optics}$}
\begin{figure}
	\centering
	\includegraphics[width=\columnwidth]{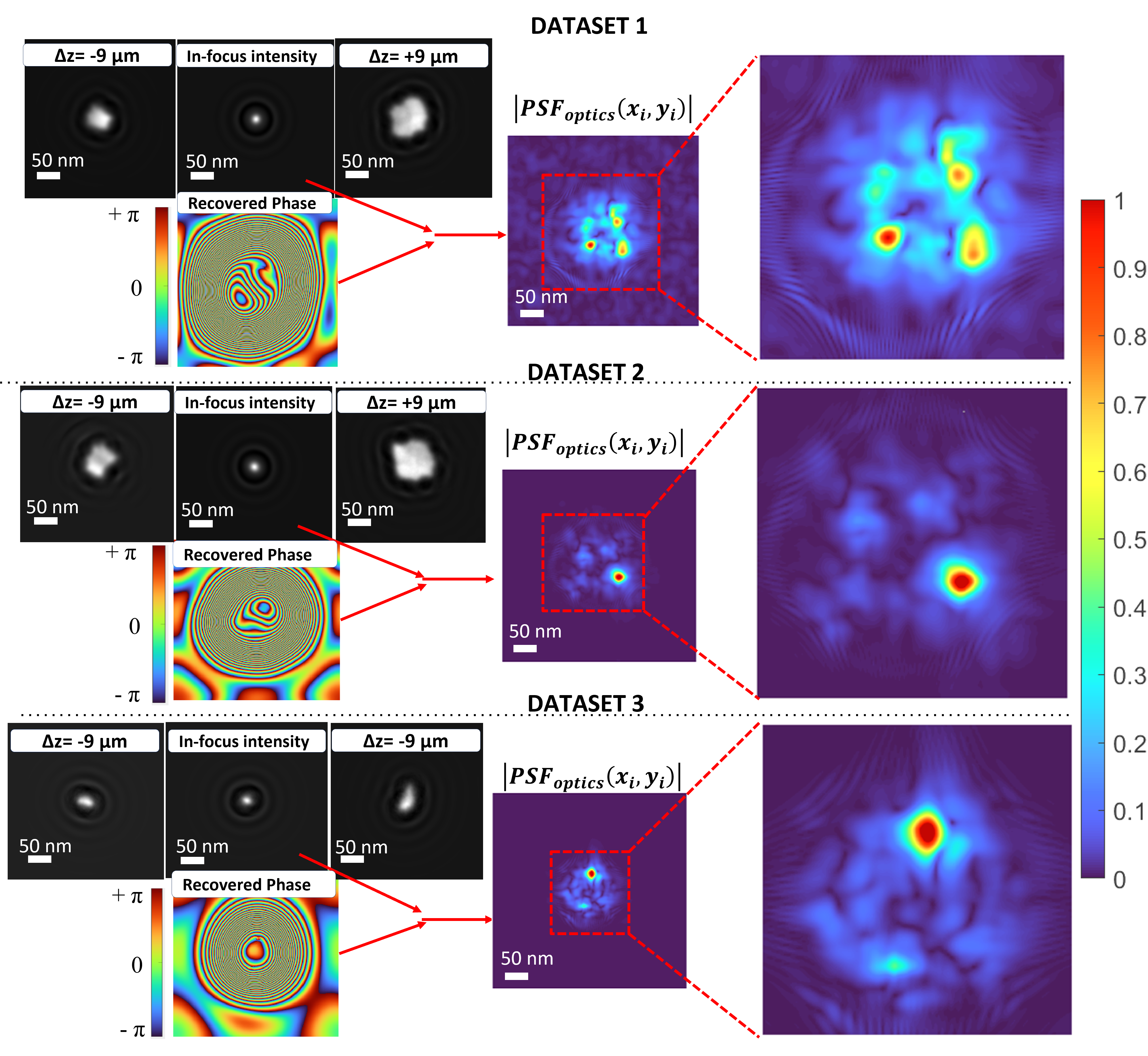}
	\caption{Visualization of the point-spread function of the probe-forming optics. For all data sets, $|PSF_{optics}|$ shows that the performance of the probe-forming optics is substantially degraded by aberrations. The zoomed-in region highlights the effect of astigmatic phase for all data sets. The zoomed-in images were contrast stretched for better visualization.}
	\label{f-68948487dacb}
\end{figure}

Since we have the phase and intensity of the probe now, we can write the complete probe wavefunction in the specimen plane (see Equation~(\ref{dfg-4d5745c8a3c6})). Hence, we can finally invert Equation~(\ref{dfg-4c44c2651550}) to calculate the coherent point-spread function of the optics ($h(x_i,y_i)\;=\;PSF_{optics} $) using Equation~(\ref{dfg-ca2060dd2e0c}). Here $\widetilde\Psi_{sp}(f_x,f_y) $ is the Fourier spectrum of the complete probe wavefunction and $\Psi_g(f_x,f_y) $ is the Fourier spectrum of the geometrical demangnification of the virtual source wavefunction ($\psi_s(x,y) $).
\let\saveeqnno\theequation
\let\savefrac\frac
\def\dispfrac{\displaystyle\savefrac}
\begin{eqnarray}
	\let\frac\dispfrac
	\gdef\theequation{8}
	\let\theHequation\theequation
	\label{dfg-4d5745c8a3c6}
	\begin{array}{@{}l}{\widetilde\psi}_{sp}(x_i,y_i)=\;\vert\psi_{sp}(x_i,y_i)\vert\cdot e^{\textstyle\{i\widetilde\varphi((x_i,y_i))\}}\end{array}
\end{eqnarray}
\global\let\theequation\saveeqnno
\addtocounter{equation}{-1}\ignorespaces 
\vskip-1.5\baselineskip 
\let\saveeqnno\theequation
\let\savefrac\frac
\def\dispfrac{\displaystyle\savefrac}
\begin{eqnarray}
	\let\frac\dispfrac
	\gdef\theequation{9}
	\let\theHequation\theequation
	\label{dfg-ca2060dd2e0c}
	\begin{array}{@{}l}\begin{array}{l}PSF_{optics}(x_i,y_i)\;=\ensuremath{\mathscr{F} }^{-1}\left\{\frac{\widetilde\Psi_{sp}(f_x,f_y)}{\Psi_g(f_x,f_y)}\right\}\\\;\end{array}\end{array}
\end{eqnarray}
\global\let\theequation\saveeqnno
\addtocounter{equation}{-1}\ignorespaces 
We show the coherent point-spread function of the probe forming optics for all the three data sets in Figure~\ref{f-68948487dacb}. A demagnification of $M\approx10 $ was used to show the results for the TESCAN MIRA3 SEM. As we added astigmatism in the data sets by varying the stigmators, the contours in the zoomed-in part can be observed which confirm presence of astigmatism in the point-spread function. The implementation of iterative phase retrieval and point-spread function visualization is open-source. We have provided our code along with data set 1 at \href{https://github.com/suryaphysics/SEM-Optics}{https://github.com/suryaphysics/SEM-Optics}.  

\begin{figure*}
	\centering
	\includegraphics[width=\textwidth]{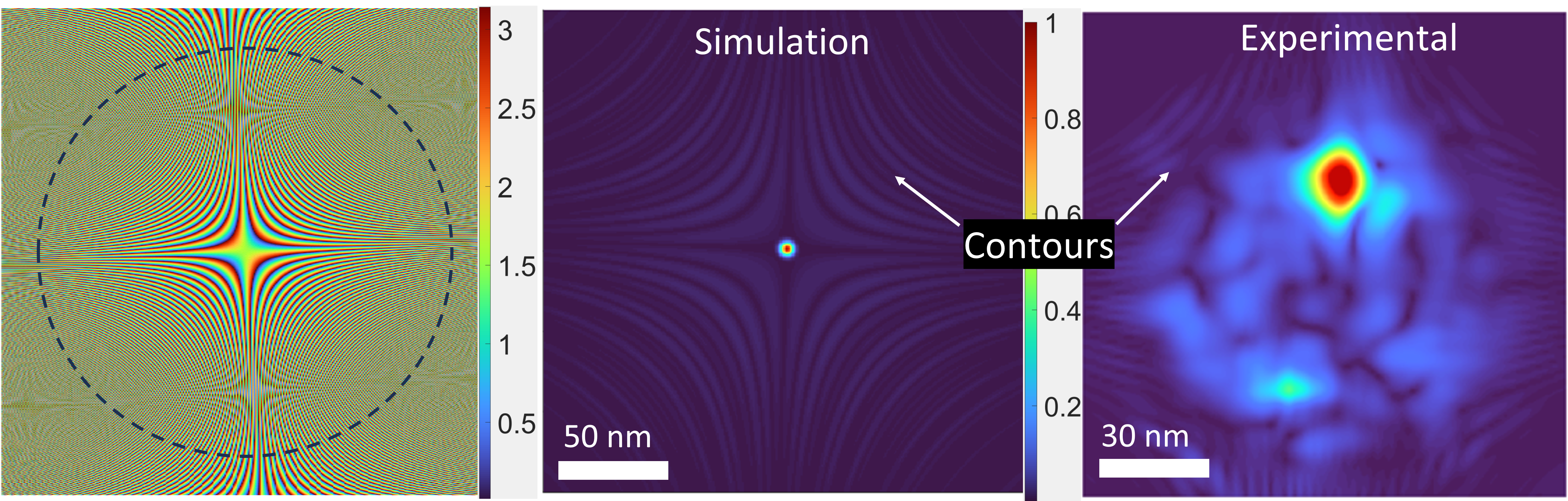}
	\caption{Qualitative comparison between a simulated astigmatic probe and the experimentally reconstructed $|PSF_{optics}|$. (Left) Aberration mask $t_{aberration}(x,y)=\pi\,\mathrm{mod}\!\left[(x^{2}-y^{2}+30xy)+\frac{1}{2},\,1\right]$, where $(x,y)$ are spatial coordinates in the lens plane normalized by the mask radius; the factor $\pi$ rescales the values to the range $[0,\pi]$. (Center) Probe intensity obtained by plane-by-plane wave propagation through the simulated optical system with a \textit{2-fold astigmatism} aberration mask. (Right) Zoomed-in experimental $|PSF_{optics}|$ for data set 3 (contrast stretched for visualization). The simulation is intended only as a qualitative sanity check to demonstrate that coherent propagation of an astigmatic wavefront naturally produces contour-like features similar to those observed experimentally; it is not a quantitative fit to the measured point-spread function.}
	\label{f-efbc2c1e0757}
\end{figure*}

To provide a qualitative physical interpretation of the contour features observed in the reconstructed $\vert PSF_{optics}\vert$, we performed a plane-by-plane propagation of the beam wavefunction \unskip~\citep{2023298:31824213} for an example SEM column shown in Figure~\ref{f-398175dbaf6c}, with a \textit{2-fold astigmatism} aberration mask (see Figure~\ref{f-efbc2c1e0757}) introduced at the objective lens plane. This configuration is analogous to deliberately introducing astigmatism through the SEM stigmators. The purpose of this simulation is not to quantitatively reproduce the experimental point-spread function, but rather to examine whether coherent propagation of an astigmatic probe naturally gives rise to contour-like features similar to those observed experimentally.

The simulated probe distribution exhibits contour features that qualitatively resemble those observed in the experimental $\vert PSF_{optics}\vert$, supporting the interpretation that these structures are consistent with coherent wavefront aberrations rather than numerical Fourier transform artifacts. However, only \textit{2-fold astigmatism} is included in the simulation, whereas the experimental probe almost certainly contains additional aberrations, including higher-order aberrations and residual optical misalignments. Consequently, differences in the peak position, central-lobe morphology, and overall probe structure are expected and should not be interpreted as disagreement between the experiment and the simplified model. A rigorous experimental--computational comparison would require quantitative estimation of the aberration coefficients from the recovered probe wavefunction, followed by simulation of the complete SEM optical system using the measured aberration coefficients. Such a quantitative validation is beyond the scope of the present work and will be addressed in our future work.

\section{Conclusion}
In summary, we demonstrate a phase-retrieval approach for the electron probe in an SEM, enabling reconstruction of the corresponding probe point-spread function ($PSF_{optics}$) from experimental data. The reconstruction is shown for three representative astigmatic beam conditions. Access to $PSF_{optics}$ may provide a useful basis for further studies in electron optics, including potential directions such as wavefront-sensing-informed aberration characterization and correction strategies. With access to calibrated column parameters and precise system measurements, a natural next step would be the quantitative estimation of aberration coefficients. This could, in turn, support the development of more comprehensive metrics for evaluating SEM optical performance, extending beyond conventional measures such as probe width alone.

\section{Acknowledgment}

Parts of this manuscript are based on material previously reported in an arXiv preprint~\citep{kamal2024point}. The work presented here incorporates a different phase-retrieval algorithm and updated final results.

\section*{Funding}
This research did not receive any specific grant from funding agencies in the public, commercial, or not-for-profit sectors.

\section*{Declaration of competing interest}
The authors declare that they have no known competing financial interests or personal relationships that could have appeared to influence the work reported in this paper.

\section*{Data availability}
We have provided our code along with one of the focal-series dataset at \href{https://github.com/suryaphysics/SEM-Optics}{https://github.com/suryaphysics/SEM-Optics}.

\printcredits
\bibliographystyle{cas-model2-names}

\bibliography{cas-refs}



\end{document}